# On application of OMP and CoSaMP algorithms for DOA estimation problem

Abhishek Aich and P. Palanisamy

*Abstract*—Remarkable properties of Compressed sensing (CS) has led researchers to utilize it in various other fields where a solution to an underdetermined system of linear equations is needed. One such application is in the area of array signal processing e.g. in signal denoising and Direction of Arrival (DOA) estimation. From the two prominent categories of CS recovery algorithms, namely convex optimization algorithms and greedy sparse approximation algorithms, we investigate the application of greedy sparse approximation algorithms to estimate DOA in the uniform linear array (ULA) environment. We conduct an empirical investigation into the behavior of the two state-of-the-art greedy algorithms: OMP and CoSaMP. This investigation takes into account the various scenarios such as varying degrees of noise level and coherency between the sources. We perform simulation to demonstrate the performances of these algorithms and give a brief analysis of the results.

## I. Introduction

Compressed sensing (CS) has been shown to be a robust paradigm to sample, process, and recover the signals which are sparse in some domain [1]. Developed recently in the last decade, it is found to be a suitable alternative to classical signal processing operations such as sampling, compression, estimation, and detection. Nyquist-Shannon's sampling theorem shows the optimal way to acquire and reconstruct analog signals from their sampled version. It states that to restore an analog signal from its discrete sampled version accurately, the sampling rate should be at least twice its bandwidth. Whereas the fundamental theorem of linear algebra for the case of discrete signals, states that the number of samples in a linear system should be greater than or equal to the length of the input signal to ensure its the accurate recovery. The samples collected from such a process are too costly- both computationally and logistically. Also, often these bounds are found to be too stringent in a situation where signals of interest are sparse, *i.e.* when these signals can be represented using a relatively small number of nonzero coefficients. Hence, to deal with such a scenario, we take help of compressive techniques like transform coding. This process finds a basis that provides sparse representation for signal of interest, thus aiming to find the most concise representation of the signal so as to trade off for an acceptable level of distortion. By sparse representation, we mean that for a signal of length $N$, we can represent it with $M << N$ nonzero coefficients.

Signal sources have been found to be sparse in spatial domain making it convenient to exploit CS to solve DOA estimation problem [2]. Direction of Arrival estimation problem is major field of study in the area of array signal processing with continuous research to eliminate the drawbacks in existing algorithms and techniques. The existing state-of-the-art algorithms such as the class of subspace-based algorithms like multiple signal classification (MUSIC), estimation of parameters by rotational invariant techniques (ESPRIT), the nonlinear least squares (NLS) method, better known as the maximum likelihood estimation method, come with certain limitations unfortunately [3]. For example, they need to have a priori knowledge of the source number, require to compute sample data covariance matrix and consecutively require a sufficiently large number of snapshots. Again, source coherency give inaccurate results as it affects the properties of covariance matrix and their time complexity is high as they involve a multiple dimensional search. Here is where CS comes into the picture to tackle these problems, prompting for further studies in the connection between array signal processing and CS theory [4] [5].

In this paper, we concentrate on the application of CS recovery based greedy algorithms out of the two categories mentioned earlier. In this context, the word "greedy" implies recovery strategies in which, at each step, we have to take a "hard" decision, generally based on some locally optimal optimization condition. The typical greedy sparse recovery approaches are basis pursuit (BP), orthogonal matching pursuit (OMP) [6], compressive sampling matching pursuit (CoSaMP) and forward backward pursuit (FBP) [7]. All these approaches make a very compelling and favorable case for solving the DOA estimation problem. The major advantages of these algorithms are low computational complexity and low time complexity for desired property recovery. The subspace-based algorithms have huge computational cost (estimation of covariance matrix and eigen decomposition) and memory cost (large number of snapshots) which creates inconvenience for real time applications. Our previous work involved using CS beamformer for improving the MUSIC algorithm by adapting the measuring matrix using an optimal bound for its dimension

Abhishek Aich is with the Department of Electronics and Communication Engineering, National Institute of Technology, Tiruchirappalli, 620 015, India. (e-mail: abhishekaich.nitt@gmail.com).

P. Palanisamy is with the Department of Electronics and Communication Engineering, National Institute of Technology, Tiruchirappalli, 620 015, India. (e-mail: palan@nitt.edu).

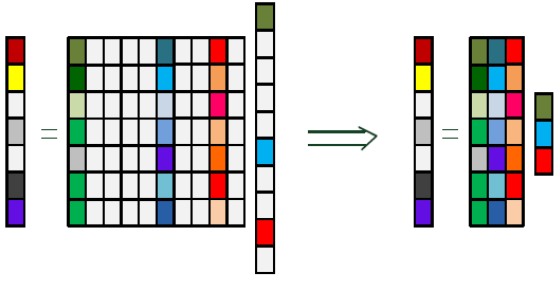

Fig.1. If the correct columns are chosen, then convert the underdetermined system into overdetermined system.

[8]. This motivates current research to implement the algorithms such as CS based algorithms to solve the DOA estimation problem, thus making them a suitable alternative for engineering practice. The next sections of the paper are organized as follows. The system model for DOA estimation problem is presented in Section II. The two greedy algorithms are explained in Section III. Section IV demonstrates the performances these algorithms and Section V concludes the work.

## II. PRELIMINARIES

### A. Data Model

Suppose $M$ narrowband source signals impinge on a uniform linear array (ULA) of $N$ omnidirectional sensors from directions $\theta_1, \theta_2, ..., \theta_M$. The output of these sensors is represented by the following model:

$$x(k) = \sum_{i=1}^{M} a(\theta_i) s_i(k) + n(k); \; k = 1, 2, ..., K$$

$K$ denotes $k^{th}$ snapshot number. $x(k) \in \mathbb{C}^N$, $s(k) \in \mathbb{C}^M$ and $n(k) \in \mathbb{C}^N$ denote the received data, source signal vector and the noise vector at snapshot time $k$ respectively. $a(\theta_i)$, $i = 1, 2, ..., M$ denotes the steering vector of the respective $i^{th}$ source. It forms the array manifold matrix $A(\theta)$ consisting of all the steering vectors $a(\theta_i)$. In matrix form, it is written as

$$X = A(\theta)S + N \quad (1)$$

where $X = [x(1), x(2), ..., x(K)]$, $S = [s(1), s(2), ..., s(K)]$ and $N = [n(1), n(1), ..., n(K)]$. The goal is to estimate $\theta_i$, $i = 1, 2, ..., M$ given $X$ and $a(\theta_i)$. Note that we will be considering only the case of single snapshot, hence the data model representation will be as

$$x = A(\theta)s + n \quad (2)$$

### B. Introduction to Compressive Sensing

An overview of CS theory can be found in [1] [9]. Let $x \in \mathbb{C}^N$ be the input signal. We need to obtain $m > M \ln(N)$ linear measurements from $x$. For this, multiply $x$ by a measurement matrix $\phi \in \mathbb{C}^{m \times N}$. This process is represented as

$$y = \phi x \quad (3)$$

with $y \in \mathbb{C}^m$. According to CS theory, for accurate reconstruction from lesser measurements, $x$ has to be sparse in some transform domain. Let's now consider the DOA data model (2) in CS environment taking the noiseless case for simplicity. Here $s$ will be the *sparse* representation of non-sparse signal $x$ in transform domain $A(\theta)$, then the overall sampling process becomes

$$y = \phi A(\theta) s \quad (4)$$

$$y = \Psi s \quad (5)$$

Here $\Psi = \phi A(\theta)$ is called the *sparsifying* or the transformation matrix. Hence, s is said to be $M$ - sparse and to reconstruct $x$ accurately, ideally $M$ measurements are required. Our aim here therefore is to recover s, given $y$. This is done by finding the sparsest solution of the following objective problem

$$\min \| s \|_0 \; s. t. \; y = \phi A s \quad (6)$$

where $\|s\|_0$ corresponds to non-zero entries in $s$. In this paper, we obtain these solutions using the greedy algorithms.

### C. Basics of Greedy algorithms

The LP technique to solve $l_1$-norm minimization problem is very effective in reconstructing desired signal. But the trade-off is that it's computationally costly. For major engineering applications like wireless communications, even the time complexity of $l_1$-norm minimization solver is prohibitive. In such cases, greedy algorithms provide a suitable alternative. By "greedy algorithm", it means an algorithm to make an optimal selection at each time locally so as to find a globally optimum solution in the end of the process. These can be broadly categorized into two strategies as "greedy" pursuits and "thresholding" algorithms.

Greedy pursuits are a set of algorithms that iteratively build up an estimate of $s$. Beginning with a zero-signal vector, these algorithms estimate a set of non-zero entries of $s$ by iteratively adding new entries which where non-zero. This selection is alternated with an estimation step in which the values for the non-zero entries are optimized. These algorithms have very less time complexity and are useful in very large data-sets. The orthogonal matching pursuit (OMP) and the forward backward pursuit (FBP) fall in this category.

Due to their ability to remove non-zero entries, the second category is called as "thresholding" algorithms. The main examples are the Compressive Sampling Matching Pursuit (CoSaMP) and the Subspace Pursuit (SP). Both CoSaMP and SP maintain track of the non-zero elements while both adding and removing entries in each iteration. At the beginning of each iteration, a sparse estimate of $s$ is used to calculate a residual error and the required indices' support is updated. Then, either the algorithms take a new estimate of this intermediate estimate of $s$, keeping it restricted to the current support set or solve a second least-squares problem restricted to this same support. We now do an analysis of the OMP and CoSaMP algorithms for DOA environment. Fig. 1 shows the pictorial representation of the concept behind greedy algorithms [10].

## III. OMP AND CoSaMP ALGORITHMS FOR DOA ESTIMATION PROBLEM

This section explains the basic idea behind the greedy algorithms – OMP and CoSaMP. A very important part of adapting these algorithms to finally obtain the set of DOAs is the plotting of angle spectrum which is common to both the above said algorithms after estimating approximation of s. We do this by using (7).

$$\mathbf{P}_s(\theta) = ||\hat{s}_\theta||^2 \,;\, \theta = \theta_1, \theta_2, ..., \theta_{N_s} \quad (7)$$

where $N_s$ being the total number of angles to be scanned. The peaks from the plot $\mathbf{P}_s(\theta)$ vs $\theta$ correspond to respective DOAs.

For the convenience of the reader, we again state the objective problem in (8) to estimate $\hat{s}$

$$\min ||s||_0 \; s.t. \; y = \Psi s \quad (8)$$

where $\Psi = \phi \mathbf{A}(\theta)$. $c$ is the current iteration number and $\Lambda_c$ is the support set at $c^{th}$ iteration. We solve (8) by following algorithms.

### A. Orthogonal Matching Pursuit

In this algorithm, the approximation for $s$ is updated in each iteration by projecting $\mathbf{y}$ orthogonally onto the columns of $\Psi$ associated with the current support set $\Lambda_c$ with $c$ denoting the current iteration. Hence it minimizes $||\mathbf{y} - \Psi s||_2$ over all $s$ with support $\Lambda_c$ and never re-selects an entry. Also, the residual at any iteration is always orthogonal to all currently selected entries. The algorithm is as follows.

---
**Algorithm III. 1:** OMP [6]

**Input**:
- $\Psi, \mathbf{y}, M$

**Output**:
- An estimate $\hat{s}$

**Procedure**:
1) Set $r_0 = \mathbf{y}$, $\Psi_0 = \emptyset$, $\Lambda_0 = \emptyset$, and an iteration counter $c = 1$
2) Find the corresponding index $\lambda_c$ of the optimization problem
$$\lambda_c = \arg\max_{j \in \{1,...,N\}} |\langle r_{c-1}, \gamma_j \rangle|$$
3) Augment the index set $\Lambda_0 = \Lambda_{c-1} \cup \{\lambda_c\}$ and the matrix of chosen atoms $\Psi_c = [\Psi_c \; \gamma_j]$
4) Solve the following optimization problem to obtain the signal vector estimate for $\Psi_c$:
$$s_c = \arg\min_s ||\Psi_c s - \mathbf{y}||_2$$
5) Calculate the new approximation ($\beta_c$) of $\mathbf{y}$ and the new residual:
$$\beta_c = \Psi_c s_c$$
$$r_c = \mathbf{y} - \beta_c$$
6) Increase $c$ by 1, and return to Step 2) if $c < M$.
7) Value of estimate $\hat{s}$ in $\lambda_j$ equals the $j^{th}$ component of $s_c$.

---

### B. Compressive Sampling Matching Pursuit

The CoSaMP algorithm applies hard thresholding by selecting the $M$ largest entries of a vector obtained by applying a pseudoinverse to $\mathbf{y}$. The columns of $\Psi$, selected for the pseudoinverse, are obtained by applying hard thresholding of magnitude $2M$ to $\Psi^*$ applied to the residual from the previous iteration and adding these indices to the support set $\Lambda_c$ from the previous iteration. Here $\Psi^*$ is the complex conjugate of $\Psi$. A major factor of CoSaMP is that it uses pseudoinverse in every iteration. Again, when computing the output vector s, CoSaMP does not need to apply another pseudoinverse as in case of OMP. Algorithm **III. 2** gives the CoSaMP algorithm.

---
**Algorithm III. 2:** CoSaMP [7]

**Input**:
- $\Psi, \mathbf{y}, M$

**Output**:
- An estimate $\hat{s}$

*Procedure*: (Loop until convergence)
1) Set $r_0 = \mathbf{y}$, $s_0 = \emptyset$, $\Lambda_0 = \emptyset$, and an iteration counter $c = 1$
2) Compute the current error
$$e = |\langle \Psi, \mathbf{y} \rangle|$$
3) Find the best $2M$ index set of $e$, $\Lambda_c = e_{2M}$
4) Update the current support set as:
$$\Lambda_c = \Lambda_{c-1} \cup \Lambda_c$$
5) Estimate the new approximation ($\beta_c$) of $\mathbf{y}$ and the new residual:
$$\beta_c = \Psi_\Lambda^\dagger \mathbf{y}$$
$$r_c = \mathbf{y} - \Psi \beta_c$$
6) Increase $c$ by 1, and set $\hat{s}_c = \beta_c$.

---

## IV. NUMERICAL ANALYSIS

In these section, we present **MATLAB** simulations to study the performance of OMP and CoSaMP algorithm in various DOA environment scenarios. Subsequent results are discussed with corresponding plots and their analysis. For all the simulations, the ULA has $N = 15$ sensors. The scanning direction grid contains $N_s = 181$ points being sampled from $-90°$ to $90°$ with $1°$ interval. Throughout this section, the noises are generated from a zero-mean complex Gaussian distribution. The number of snapshot is $K = 1$, hence the treated as a single measurement vector (SMV) problem.

**Simulation 1:** In this example, there are three non-coherent sources with respective directions as $\theta_1 = -60°$, $\theta_2 = 0°$ and $\theta_3 = 40°$. The SNR is set to be 0 dB to create a noisy environment. The performances of the algorithms are shown in Fig.2. We can observe from the plot that the OMP algorithm correctly resolves all the sources under the given scenario with better resolution. The plot of CoSaMP spectrum however gives poorer results and is not able to detect all the sources properly. This is mainly because it needs more measurements than what OMP requires to correctly detect the DOA support set. This was confirmed by taking a larger number of measurements.

**Simulation 2:** Setting the same DOA environment as the first example, we have a source from $\theta_1 = -60°$ and two coherent sources with directions $\theta_2 = 0°$ and $\theta_3 = 40°$. The SNR remains unchanged. The performances of the algorithms are now observed in this partially coherent source environment in Fig.3. We can observe from the plot that the OMP algorithm accurately resolves all the sources under the partially coherent source scenario without false peaks. However, CoSaMP fails to resolve all the signals while generating false peaks. Thus, OMP algorithm works well even here, given the coherency of the sources. These two examples show an empirical advantage of OMP over CoSaMP in terms of performances.

We now do the RMSE *vs* SNR analysis to observe the performance for 1000 trials.

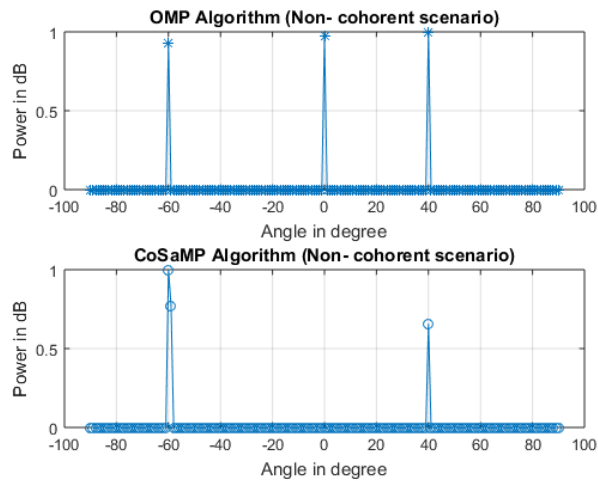

Fig. 2. Plot for Simulation 1 (N = 15, M = 3, SNR= 0 dB)

**Simulation 3:** This simulation considers two sources with DOAs as $\theta_1 = -60°$ and $\theta_2 = 60°$. we compare the algorithms with respect to root mean square error (RMSE) *vs* SNR (dB). 1000 independent Monte Carlo experiments are performed. It is observed from Fig. 4 that the OMP algorithm achieves a much better estimation performance.

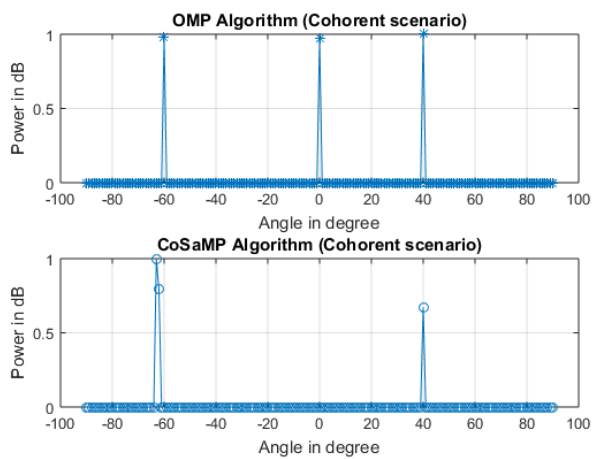

Fig. 3. Plot for Simulation 2 (N = 15, M = 3, SNR= 0 dB)

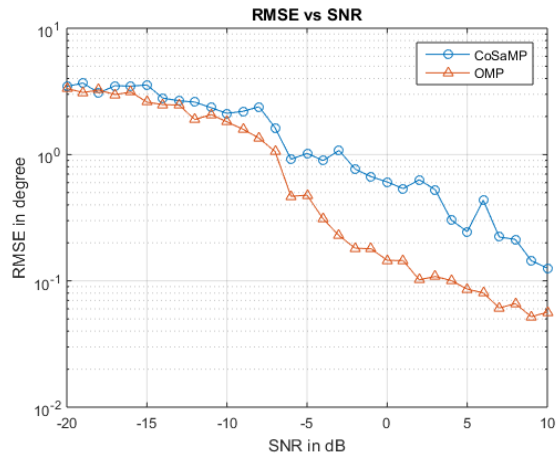

Fig. 4. Plot for Simulation 3 (N = 15, M = 2, Trials = 1000)

## V. CONCLUSION

In this paper, we provided an overview of the application of the OMP and CoSaMP algorithm to the DOA estimation problem by modelling it as a standard CS recovery problem. The major advantage of these algorithms is that they don't require any eigen value decomposition and work well with single snapshot. This is highly desirable for practical engineering applications such as dynamic tracking of a vehicle. It can be fairly concluded that greedy algorithms are a favorable candidate for DOA estimation problems as they are fast and have high resolution. These algorithms remain at the forefront of active CS research and thus, provide a strong alternative tool for wide range of array signal processing applications. We will further work on the performance improvement of these algorithms specifically by designing an adaptive dictionary to suit the DOA estimation problem.